# Innovative Modular Design and Kinematic Approach based on Screw Theory for Triple Scissors Links Deployable Space Antenna Mechanism

Mamoon Aamir[a], Mariyam Sattar[a*1], Naveed Ur Rehman Junejo[b*2], Aqsa Zafar Abbasi[c]

[a]Department of Aeronautics and Astronautics, Institute of Space Technology, Islamabad, Pakistan
[b]The University of Lahore, Lahore Campus, Pakistan.
[c]Department of Applied Mathematics and Statistics, Institute of Space Technology, Islamabad, Pakistan

[*1]Corresponding Author 1: mariya98975@gmail.com
[*2]Corresponding Author 2: naveed.rehman@dce.uol.edu.pk

**Abstract**
This paper presents the geometry design and analysis of a novel triple scissors links deployable antenna mechanism (TSDAM) to deal with the problems of large-aperture and high-precision space antennas for deep-space communication and Earth observation. This mechanism has only one degree of freedom (DoF) and thus makes for efficient and reliable deployment without loss of structural integrity. It employed a systematic design approach starting from a triple scissors links modular unit to a 25m aperture assembly. Different configurations constituting variable numbers of modular units were analyzed in SolidWorks to identify the deployable mechanism with lowest deformation. While the 24-units configuration offered superior stowage compactness, it exhibited higher deformation (0.01437mm), confirming the 12-units configuration as the optimal balance between structural stability and deployment efficiency. Screw theory was employed to analyze the kinematic properties, and numerical simulations were performed in MATLAB and SolidWorks. The deployable space antenna showed transition from stowed to fully deployed state in just 53 seconds with high stability throughout the deployment process. The TSDAM attained a storage ratio of up to 15.3 for height and volume with 0.01048mm of deformation for a 12-unit's configuration. Mesh convergence analysis proved the consistency of the simulation results for 415314 tetrahedral-shaped elements. The virtual experiments in SolidWorks verified the analytical Screw theory based model and ensured that the design was smooth and flexible for deployment in operational conditions. The research establishes a robust design framework for future deployable antennas, offering enhanced performance, simplified structure, and improved reliability

**Keywords-** DoF; Kinematic Analysis; Triple scissors link deployable antenna mechanism; Screw Theory; Virtual experiments; Storage ratio

## 1. Introduction

Deployable mechanisms are very extensively used to construct large space structures. Among these, the deployable flexible cable net antenna are the most in-demand mechanisms in aerospace applications due to high storability and lightweight [1, 2]. Extensive research in aerospace science and technology has led to innovative designs and analysis methods to further improve the deployable antenna mechanisms. However, these mechanisms still require further improvements to meet the needs of different communications and data transmission missions [1, 3, 4]. Deployable space mechanisms that compactly fold for transport and expand in orbit, are widely used in satellite and spacecraft technology. The large diameter antennas generally develop the deployable form for use because the launcher construction spaces are usually small to deploy big sizes [5-8]. These structures enhance loading capacity and adaptability but their complex, multi-joint, closed-loop mechanisms pose challenges in kinematic and dynamic analysis [9, 10]. As aerospace research and technology advance quickly to meet the demands of various space activities, the gain needs of space antennas are rising quickly [1, 11]. Simple and effective method to improve the gain of antenna is the increment in diameter, so large antennas are often deployable [5, 7, 8]. The deployable configuration approach has, therefore, gained significant attention in recent years [12-14]. Furthermore, a deployable mechanism that folds rapidly with excellent accuracy and reliability is currently the focus of extensive study worldwide [15]. Scissors grids are a category of deployable structures in the form of articulated bars. Deployable truss mechanisms offer exceptional functioning in stowage space, structure stability, high surface accuracy and stiffness [16-19]. Numerous researchers have considered different arrangements of modular structures considering the basic deployable units that constitute the large deployable mechanisms. Tetrahedral truss, Fig. 1 (a), ring truss, Fig. 1 (b) and radial rib deployable antennas are the most common space deployable mechanisms in orbit [14, 20-23]. The ring truss deployable antenna is ideal for large diameter space applications due to its light weight and high folding ratio [14]. Moreover, the mass of deployable space mechanisms does not grow proportionately with an increase in diameter [24-27]. The structural designs of truss mechanisms for space deployable antenna have been extensively discussed [28-30]. Xu et al., examined the tetrahedral deployable truss [31] while, Gao et al., proposed a symmetric dodecahedron mechanism that formed the basis for a unique deployable coupling system [32]. A deployable flat-panel triangular pyramid unit with one DoF was examined by Ding et al., [33]. Researchers also created a chain of deployable configurations by joining spatial single-loop links [34-36], and employed the scissors mechanism to join two distinct types of double-ring deployable trusses [37, 38]. However, Li et al., [39] considered the deployable units with spherical scissors mechanisms and Tang et al., [40] introduced the bifurcated 8 bar linkages that formed the base of a deployable mechanism.

Most of the proposed antenna have multiple DoF, which require synchronous joints that increase complexity and reduce deployment reliability. Deployable structures are closed-loop coupled mechanisms with joints and rod assemblies presenting challenging dynamic and kinematic studies. Wei et al., performed geometry analysis of a wide range of deployable closed-loop mechanisms [41, 42]. The process is generalized to obtain the mobility features of the assembly. When the deployable antenna mechanism reaches a predefined orbit, it must be deployed carefully. The deployment must be gentle to prevent impacts that could affect accuracy. Sattar et al., [43] applied Newton's method to analyze the kinematics and dynamics of a double scissors links truss deployable antenna, Fig. 1 (c), that presented high storage efficiency. Ding et al. [33] investigated the kinematic properties of polyhedral applying numerical analysis. Tian et al. [44] considered the coordinate transformation method to create a kinematic model of a deployable mechanism. Using the displacement vector approach, Xu et al. [45] constructed the kinematic model of a space manipulator. Sun et al. [46] examined the grid computing technique to determine the kinematics of scissors mechanisms. The kinematics of the mechanism in the screw coordinates was examined by Zhao et al. [47], while the Finite-Step-Integration method was introduced by Han et al. [48] to analyze the forward kinematics of parallel mechanisms. Liu et al. [49] and Qi et al. [50] used the screw theory and generalized coordinates, respectively, to investigate the kinematics of the space capture mechanism. In addition, researchers have applied the generalized, absolute nodal coordinates and Lagrange approach to develop the dynamic equations of the deployable space mechanisms [26, 51-53]. Bai et al. [54] examined the dynamics of the antenna system taking into account the changes in the side gaps, while Zhao et al., [55], analyzed the dynamic properties of a very large one-hundred-meter aperture ring antenna.

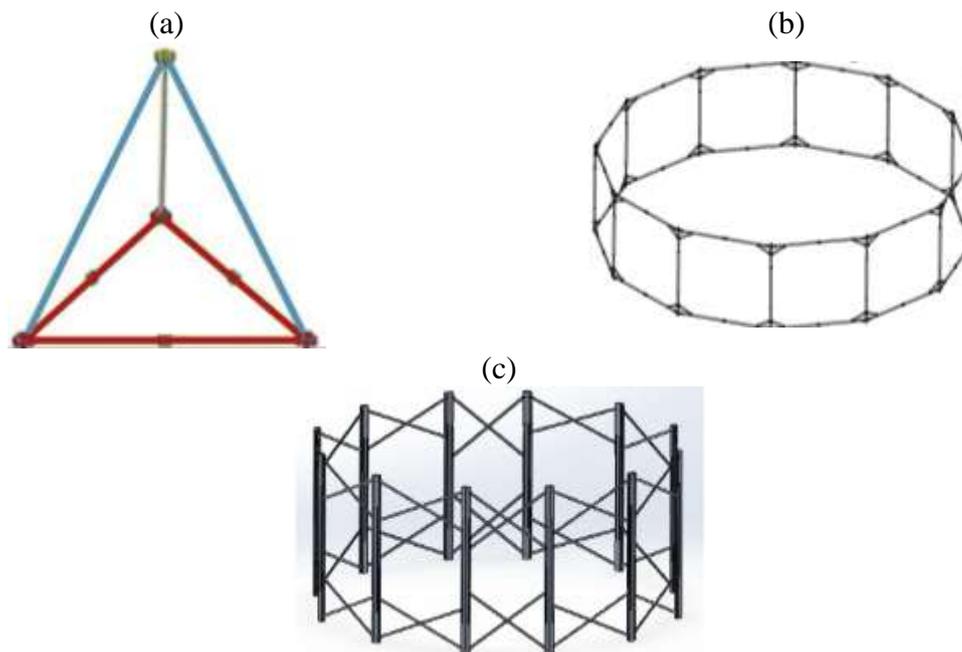

**Fig. 1.** Different deployable antenna configurations (a) Tetrahedral truss [22] (b) Single ring truss deployable antenna [14] (c) Double Scissors links truss deployable mechanism [43]

While extensive research has explored kinematic properties in the space mechanisms, few generalized techniques address the complexities of TSDAM kinematics. It poses the challenges in applying the theoretical models to real-life applications due to the factors such as variation in length of the links, frequency, and stiffness. This study addresses this gap by utilizing the screw theory method. Most existing designs involve multi-loop trusses with high DoF, leading to unreliable configurations. Synchronous joints often used to achieve these configurations add complexity and reduce system reliability. This paper proposes a novel approach by analyzing the feasibility of a TSDAM with 1-DoF. The research integrates the advanced simulations with analytical methods to bridge the gap between theory and practice thus offering a reliable design method for deployable structures. The TSDAM configuration is designed using Aluminum Alloy 7075-T6 due to its high strength-to-weight ratio (UTS ~572 MPa), excellent stiffness (E ~71.7 GPa), and proven endurance in harsh aerospace dynamic conditions. The rest of this paper is organized as follows: Section 2 discusses the problem statement depicting the main objectives. A class of novel TSDAM to produce antenna of 25m aperture is proposed in Section 3. Section 4 deals with its design, modeling and geometrical relationships. Section 5 simulates the deformation response of deployable antenna mechanism with different installations. Section 6 computes the DoF by screw constraint topological graphs and screw theory. The analytical kinematics analysis through the derivation of constraint relations based on screw theory is detailed in Section 7. The validation of theoretical and virtual experiments results along with comparative study is given in Section 8. Section 9 concludes the work by summarizing the principal results and their implications.

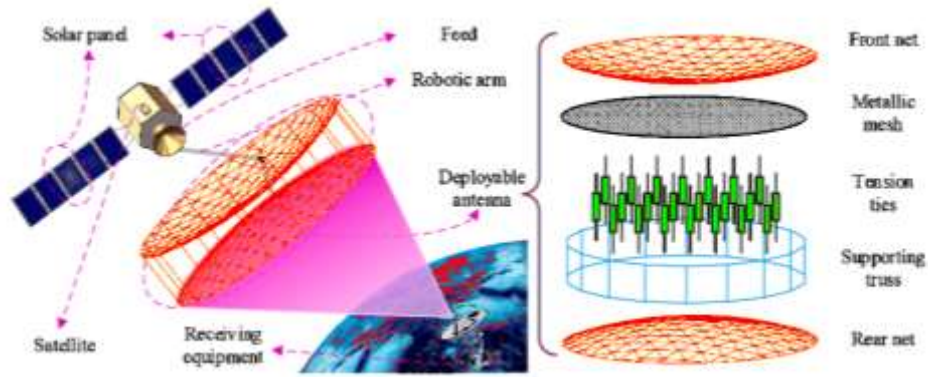

**Fig. 2.** Main components of satellite space antenna [56]

## 2. Problem Statement

The novel triple scissors design proposed in this research presents an efficient, reliable, and lightweight solution for large aperture deployment in space. Fig. 3 shows the comprehensive methodology frame work that explains the flow of design, analytical and simulation procedures to investigate and validate the integrity of the novel mechanism for space antenna. Existing large-aperture, 25m to 30m, deployable antenna suffer from shortcomings based on complexity, weight, and deployment precision that limits their usage on the high-performance missions. Likewise, traditional mechanisms apply complicated multi-step deployment processes that often lead to inefficiencies and failure risks in space environments. However, the space antenna truss mechanism designed by utilizing the proposed simplified novel triple scissors links modular units can resolve the problems of large aperture associated with the reliability of deploying unit. This research aims to study the feasibility and performance of a 25m aperture antenna formed from the novel triple scissors deployable modular mechanisms. In order to maximize modeling precision and applicability, the deployable mechanism utilizes Aluminum 7075-T6 due to its superior fatigue durability and weight efficiency. The key focus is on the kinematics and structural integrity of the system. Screw theory is applied to study the analytical kinematics because it offers an efficient and mathematically robust basis to analyze both the translation and rotation in complex as well as novel mechanical systems. A thorough investigation is performed to validate the deployment sequence, forces, and motion of the mechanism through virtual experiments in SolidWorks and MATLAB, ensuring its suitability for the challenging space exploratory missions.

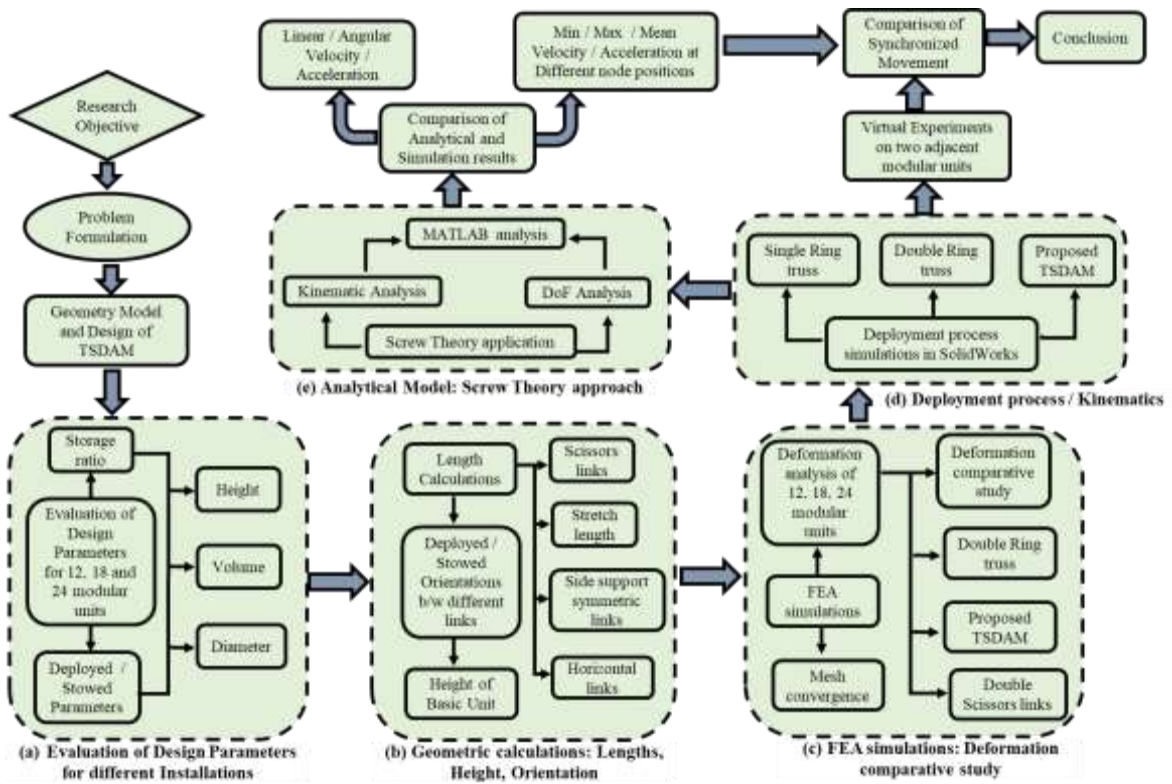

**Fig. 3.** Complete methodology flowchart for the TSDAM study

## 3. Proposed Mechanism for Design of Deployable Modular Unit

The basic modular units in antenna are interconnected to form a circular deployable mechanism. It is developed in a systematic manner to ensure a clear progression from a simple configuration to the final design. Fig. 4 shows step by step development of module from basic scissors links and their integration into the finalized basic deployable unit. As shown in Fig. 4 (a), initially, a single scissors link was considered to form the basic building block of the modular mechanism.

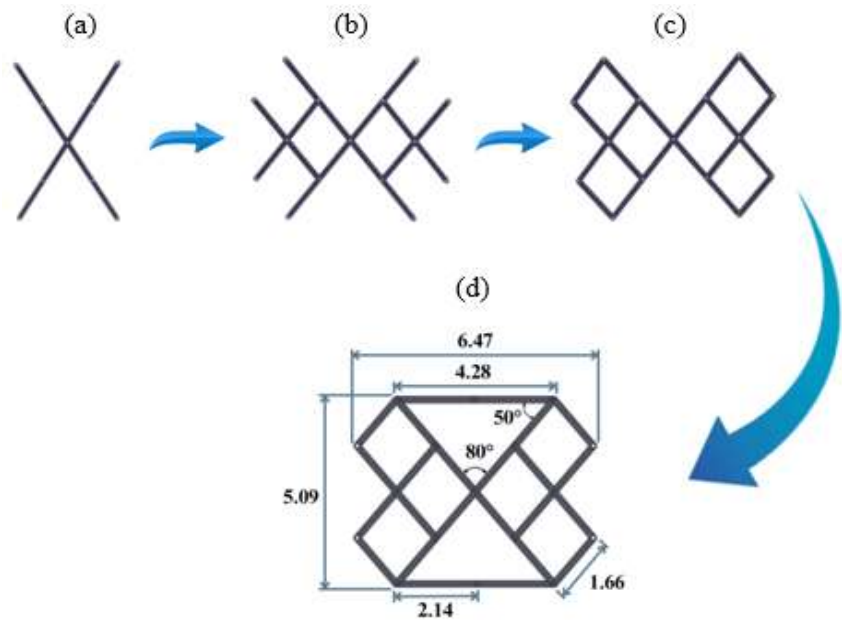

**Fig. 4.** Development of the modular unit for novel space antenna mechanism (a) single scissor link, (b) symmetric side scissors links, (c) supporting links, and (d) horizontal links.

Next, two additional scissor links were added symmetrically on each side of the initial link, extending the overall span and establishing the core framework, Fig. 4 (b). To enhance structural support and stability, supporting links were introduced between the scissor units, Fig. 4 (c). These links played a crucial role in maintaining the alignment and rigidity of the expanding structure. Finally, four horizontal links were integrated across the configuration, Fig. 4 (d). These horizontal members at the top and bottom of the triple scissors links assembly ensured uniform load distribution and fix the stretched length of the modular unit. This step-by-step approach allowed for an organized assembly, ensuring that each stage contributed to the overall stability and functionality of the antenna. It can be observed that the fully deployed modular unit is constrained at an angle of 80° between the scissors links, while the top and bottom links are retrained at an angle of 50°. This novel unit consisting of triple scissors links mechanism is optimized to provide high storage ratio that makes it highly appropriate for large-aperture space antennas with good structural integrity. The 25m aperture offers a large surface for effective signal capture and transmission in space communication.
Fig. 5 contains the detailed stowed structure of TSDAM. It can be observed that fully stowed configuration exhibits an angle of 12.54° between the scissors and top and bottom links.

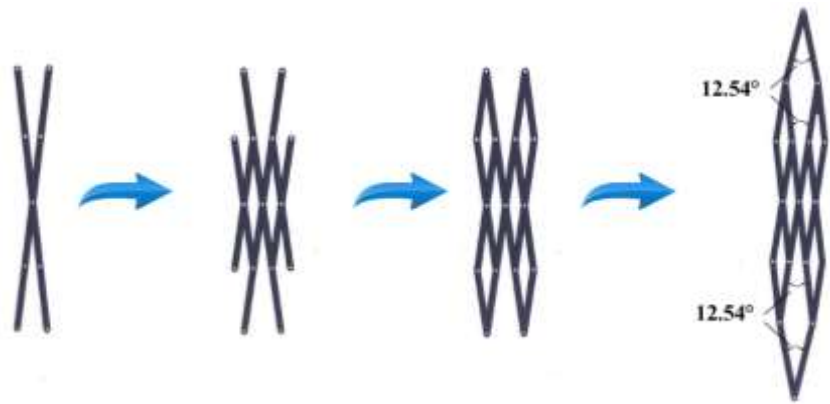

**Fig. 5.** Stowed configuration for modular unit of TSDAM

## 4. Geometry model and design of novel TSDAM

Fig. 6 shows that the basic deployable unit is designed as four horizontal links and ten diagonal links. This modular assembly was configured for an antenna of 25m aperture. A detailed comparative analysis of deformation patterns (Fig. 11) and storage ratios (Table 2) is performed for a 25m aperture antenna formed from 12, 18 and 24 deployable modular units.

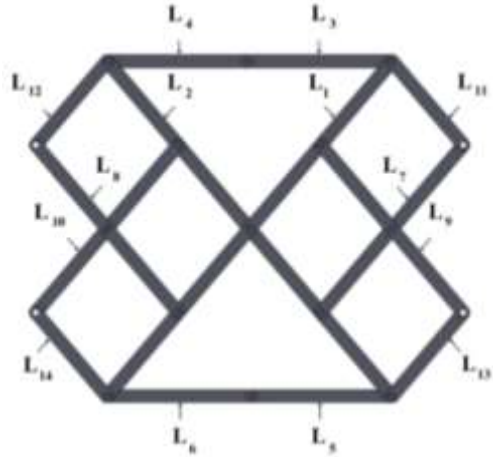

**Fig. 6.** Geometric layout of basic deployable triple scissors links modular unit

The space antenna truss assembly consisting of 12 deployable modular units is selected for further research due to it optimal structural integrity and performance. Therefore, the stretched length of the basic module for 12 units antenna assembly is calculated to be 6.47m. To achieve the best possible deployment, the angle between the two-diagonal scissors links, labeled as $L_1$ and $L_2$ was constrained at 80° for the fully deployed configuration. Based on the principles of geometry and trigonometric relationships, the precise dimensions of all the links were systematically computed for the triple scissors link modular unit.

The height of the basic unit was set to be 5.09m. Now, in order to calculate the length of link $L_3$ we can form a right triangle and determine the unknown length.

$$\tan(40°) = \frac{L_3}{2.545} = 2.14m \quad (1)$$

Since, the antenna is symmetrical; the lengths $L_3$, $L_4$, $L_5$, and $L_6$ are equal. To calculate the lengths of diagonal links $L_1$ and $L_2$, again we considered a right triangle and set the links $L_1$ as hypotenuse. The angle between links $L_1$ and $L_3$ is 50°. The mathematical formulation given in equation (2) is applied to calculate the length of links $L_1$ and $L_2$.

$$\sin(50°) = \frac{5.09}{L_1} = 6.645m = L_2 \quad (2)$$

The length of diagonal links $L_7$, $L_8$, $L_9$ and $L_{10}$ were half of length for link $L_1$, equation (3).

$$L_7 = L_8 = L_9 = L_{10} = 3.323m \quad (3)$$

Similarly, the length of diagonal links $L_{11}$, $L_{12}$, $L_{13}$ and $L_{14}$ were half of the link $L_7$.

$$L_{11} = L_{12} = L_{13} = L_{14} = 1.662m \quad (4)$$

The dimensions calculated for 14 different links that constitute the triple scissors links deployable modular unit are given in Table 1.

**Table 1.** Length (L in meters) for all links of the modular unit

| Link | $L_1$ | $L_2$ | $L_3$ | $L_4$ | $L_5$ | $L_6$ | $L_7$ | $L_8$ | $L_9$ | $L_{10}$ | $L_{11}$ | $L_{12}$ | $L_{13}$ | $L_{14}$ |
|---|---|---|---|---|---|---|---|---|---|---|---|---|---|---|
| L (m) | 6.64 | 6.64 | 2.14 | 2.14 | 2.14 | 2.14 | 3.32 | 3.32 | 3.32 | 3.32 | 1.66 | 1.66 | 1.66 | 1.66 |

Calculations related to stretch length, deployed / stowed height, deployed / stowed diameter, deployed / stowed volume, and storage ratios for antenna with 12, 18 and 24 modular units are performed while fixing the aperture to 25m. The structure can be configured with various numbers of modular units depending on the required aperture size as shown in Table 2. Although the 18 and 24 units modular assemblies offered a good balance between compact stowage and efficient deployment but their deformation analyses (Fig. 11 and Table 5) showed high levels of structural instability. While, the antenna assembly comprising of 12 modular units provide a larger deployed height (5.122m) but it maintains a sufficient stable deployed diameter of 25m. Moreover, it shows the minimum deformation of 0.01048mm without compromising the structural integrity that makes it more suitable for our specific mission requirements. There are two variables that are independent in the suggested design. The first factor is the antenna's diameter, which is dictated by its intended use. Creating a big deployable antenna having a diameter between 20m to 30m is the aim of this study. Therefore, for the purpose of this investigation the diameter of deployable antenna is selected to be 25 m. The second independent variable, i.e., number of modular units establish the height of each link, the length of a completely deployed unit, the length of movable slots, and the diagonals. A full-scale assembly comprising of 12 modular units and an aperture of 25m was developed in SolidWorks. This 3D assembly provided a detailed representation of each component i.e., central links, and interconnected scissor mechanisms.

**Table 2.** Evaluation of design parameters for 12, 18 and 24 modular units antenna assembly

| Units | 12 | 18 | 24 |
|---|---|---|---|
| **Stretched Length (m)** | 6.470 | 4.34 | 3.26 |
| **Deployed Height (m)** | 5.09 | 3.436 | 2.581 |
| **Stowed Height (m)** | 11.010 | 7.386 | 5.548 |
| **Deployed Diameter (m)** | 25 | 25 | 25 |
| **Stowed Diameter (m)** | 3.246 | 2.176 | 1.634 |
| **Deployed Volume (m³)** | 2400.584 | 724.61 | 307.109 |
| **Stowed Volume (m³)** | 86.979 | 26.211 | 11.114 |
| **Storage Ratio (Diameter)** | 7.702 | 11.5 | 15.3 |
| **Storage Ratio (Height)** | 0.465 | 0.465 | 0.465 |
| **Storage Ratio (Volume)** | 27.6 | 27.6 | 27.6 |

The stowed and deployed angles can be calculated for length of rods (L), angle between rods (θ), overall width (W) and height (H). The cosine laws are applied to compute the deployed and stowed states of modular mechanisms as given in equation (5) and (6) respectively.

$$W = L\sqrt{2(1 + \cos(\theta_1))} \quad (5)$$

$$H = L\sqrt{2(1 + \cos(\theta_2))} \quad (6)$$

Where $\theta_1$ and $\theta_2$ in the above equations are the deployed and stowed angles respectively. Substituting W = 5.09m and L = 3.32m in equation (1) verifies $\theta_1$ = 79.90° ≈ 80°.

## 4.1 Deployment Process Simulation in SolidWorks

Fig. 7 shows the deployment progression of interconnected units of the 12-module deployable structure, where each unit is joined to the next. The stowed to fully deployed state for TSDAM is achieved in 53s and the complete cycle is accomplished in 102s. The comparative analysis for the proposed 25m aperture TSDAM is provided in Table 3.

**Table 3.** Comparative analysis of deployment process time of deployable antenna in seconds

| Antenna Mechanism | Aperture (m) | No. of units | Deployment Process time (sec) | | Complete cycle (sec) |
|---|---|---|---|---|---|
| | | | Intermediate | Complete deployed | |
| **Triple scissor** | 25 | 12 | 26 | 53 | 102 |
| **Single ring truss** | 25 | 24 | 50 | 140 | -- |
| **Double ring truss** | 5 | 18 | 60 | 102 | -- |

It can observed that the deployment process of proposed deployable mechanism is comparatively far more efficient than the single ring [57] and double ring [28] truss antenna that required 140s and 102s, respectively, to achieve fully deployed state.

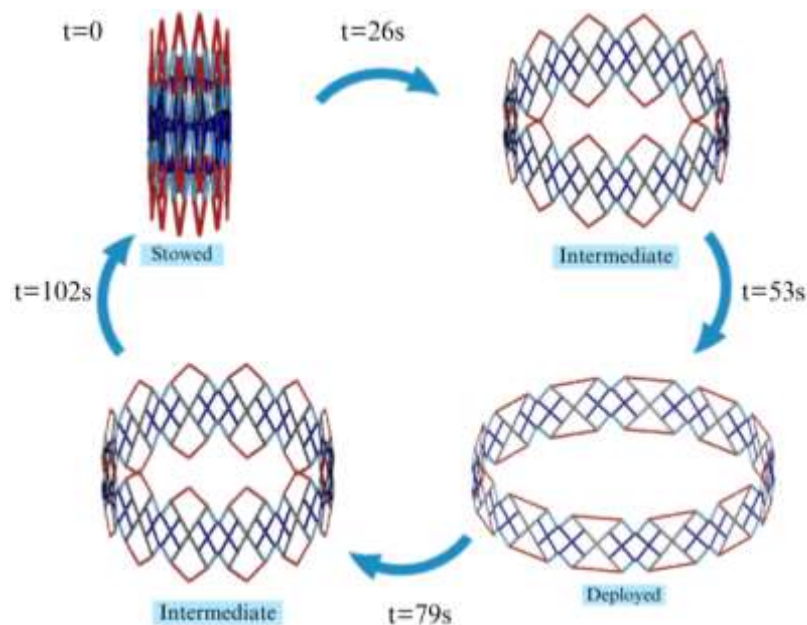

**Fig. 7.** Cycle of operation for TSDAM assembly i.e., stowed – intermediate – fully deployed – internmediate – stowed state of 12 modular units antenna

Fig. 8 provides a zoomed-in view of the joint, highlighting its design and role in enabling smooth deployment and structural cohesion.

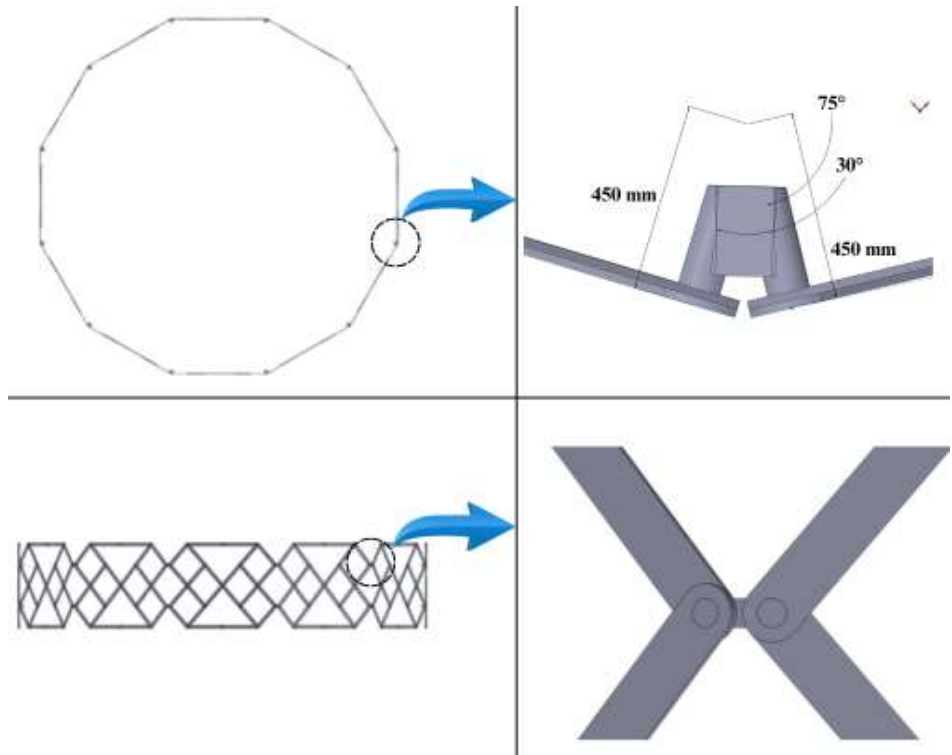

**Fig. 8.** Zoomed – in view of the joints involved in smooth deployment of TSDAM

## 5. Finite element simulations to investigate the deformation with different installations
## 5.1 Mesh and Mesh convergence study

Fig. 9 shows the meshed geometry of the 12 modular units and 25m aperture TSDAM. The precise mesh of the deployable antenna mechanism ensures accuracy of results in the virtual experiments. The meshing performed in SolidWorks employed tetrahedrons to discretize the antenna assembly. The initial mesh consisted of 327760 with orthogonal quality of 0.26698 that followed subsequent mesh refinements. For reliable analysis, the quality of the mesh was assessed considering the factors such as the skewness of the elements and their aspect ratio.

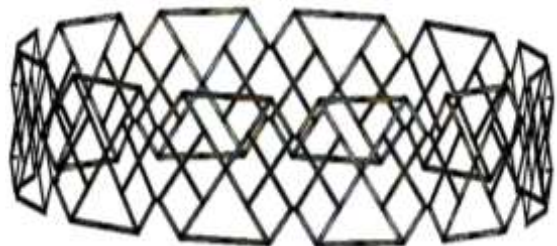

**Fig. 9.** Meshed geometry for 12 modular units TSDAM assembly

The results of the mesh convergence study and the change in the deformation after subsequent mesh refinements are shown in Fig. 10 and Table 4 respectively. To find the ideal mesh density, indicating mesh-independent outcomes, a total of nine iterations were performed. The final mesh consisting of 415,314 tetrahedral-shaped elements provided the best balance between computation speed and solution accuracy.

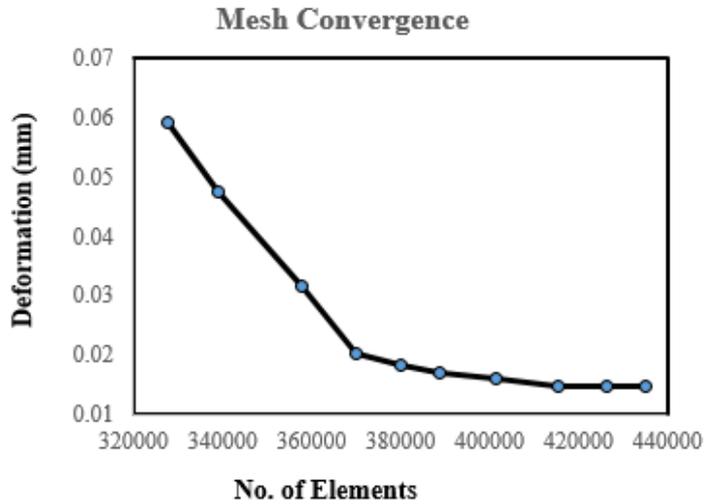

| Sr. # | Elements number | Deformation |
|---|---|---|
| 1 | 327760 | 0.0592 |
| 2 | 339156 | 0.0476 |
| 3 | 357701 | 0.0317 |
| 4 | 369960 | 0.0201 |
| 5 | 380017 | 0.0183 |
| 6 | 388628 | 0.0171 |
| 7 | 401597 | 0.0159 |
| 8 | 415314 | 0.0148 |
| 9 | 426160 | 0.0148 |

Table 4. Mesh Convergence Study

**Fig. 10.** Mesh convergence analysis based on number of elements against deformation

## 5.2 Deformation Analysis

In the design of space-deployable antennas, deformation analysis is performed to test and analyze their reliability and functionality against different conditions in space operations.

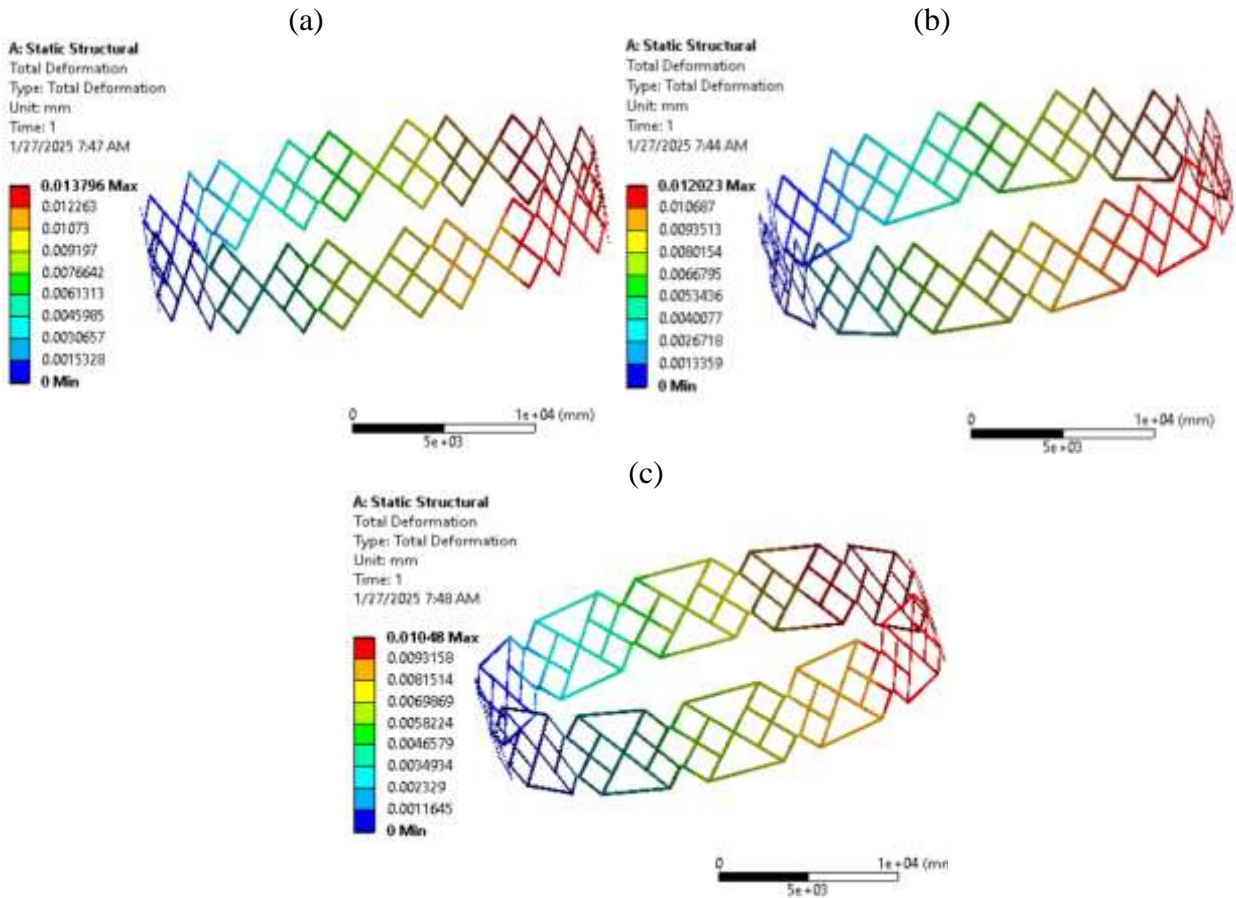

**Fig. 11.** FE analysis for deformation response of TSDAM with 12 modular units (a) without links (b) with lower horizontal links (c) with both upper and lower horizontal links

Finite element analyses were performed based on the mechanical properties of Aluminum 7075-T6. This aerospace-grade material with Young's Modulus of 71.7 GPa and yield strength of 503 MPa allows accurate assessment of stress patterns, deformation modes, and stability of structures under simulated orbital loading conditions. Fig. 11 shows the responses for maximum deformation obtained through virtual experiments on TSDAM with 12 modular units and the extended numerical results for different installations are presented in Table 5. Fig 11 (a) – Fig. 11 (c) show the antenna assembly without links, with lower horizontal link and with both upper and lower horizontal links, respectively. When the TSDAM assembly is designed, the links installed on the lower and upper side of the mechanism are entirely the same. Although, the addition of the horizontal links add some extra mass to the assembly but the overall mechanism becomes more diversified with an increase in the mechanism strength. The maximum deformation for 12 units TSDAM is observed for assembly devoid of any horizontal links, i.e., 0.01379mm. Installation of the lower horizontal links in TSDAM reduces it to 0.01202mm whereas the least occurs in the assembly with both top and bottom horizontal links attached, i.e., 0.01048mm. Table 5 shows that the 12 units antenna (with top and bottom links) presents the best structural integrity with least deformation for further kinematics analysis.

**Table 5.** Analysis of Maximum Deformation: Comparison of Proposed Model vs. Literature

| Analysis Parameters | Deployable Antenna Configuration | | | | |
| --- | --- | --- | --- | --- | --- |
| | Triple Scissors links antenna (TSDAM - Proposed) | | | Double Scissors [49] | Double Ring Truss [50] |
| **No. of deployable modular units** | 12 | 18 | 24 | 12 | 18 |
| **Deployable antenna aperture Dia. (m)** | 25 | 25 | 25 | 25 | 5 |
| **Max. deformation without links (mm)** | 0.01379 | 0.02107 | 0.02907 | - | - |
| **Total mass without links (kg)** | 2.1035 | 3.0701 | 4.0190 | - | - |
| **Max. deformation with lower links (mm)** | 0.01202 | 0.01680 | 0.02190 | - | - |
| **Total mass with lower links (kg)** | 2.5106 | 3.6493 | 4.7881 | - | - |
| **Max. deformation of entire antenna (mm)** | **0.01048** | 0.01090 | 0.01437 | **0.62815** | **0.073066** |
| **Total mass with top and bottom links (kg)** | 2.9178 | 4.2349 | 5.5571 | - | - |

Table 5 shows the deformation analysis of deployable antenna with different installations i.e., 12, 18 and 24 modular units assembly. It is analyzed earlier in section 4 that antenna with 24 modular units shows best balance for compact stowage and deployment efficiency. But the Table 5 presenting the analysis for structural deformation shows a stark compromise. The same antenna mechanism will exhibit very little deformation for a 12-unit antenna compared to a very critical level of deformation for a 24-unit antenna assembly. The comparative study reveal that the 12-unit antenna, due to least deformation, is the optimal solution for the technical applications. The reduced strain will assure mechanical robustness and durability, making it a preferable alternative for functions requiring stronger structural stability.

## 6. Application of Screw Theory: Degree of Freedom (DoF) Analysis

Fig. 12 shows the screw constraint topology graph for analyzing the DoF for TSDAM modular unit. The origin of the coordinate system (O – XYZ) is defined at node O. The X-axis in this case is parallel to the connection between nodes E and Q, the Z-axis is drawn vertically upward, and the Y-axis adheres to the right-hand rule. The nodes are represented in the form of circles. Lines are used to express the kinematic pairs as well as the constraints imposed by the nodes, where $S_1$, $S_2$, $S_3$, and so on represent the constraint names. These constraints appear as directed edges outlining the interactions between several components.

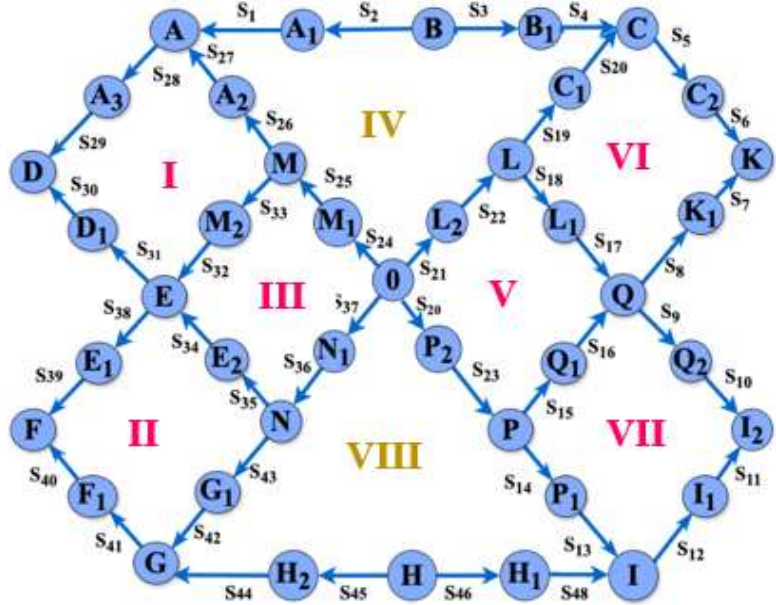

**Fig. 12.** Screw constraint topology graph for triple scissors links deployable modular unit

The relevant mathematical expressions are mentioned in equation (7) to equation (9).

$$r = \left(n + Lsin\left(\frac{\theta}{2}\right) \quad 0 \quad Lcos\left(\frac{\theta}{2}\right)\right) \tag{7}$$

The direction of the axis is given by equation (8).

$$\mathbb{S} = (0 \quad 1 \quad 0) \tag{8}$$

According to the screw theory the motion of the joints between two scissors links (S) is given by mathematical formulation in equation (9).

$$S = \left(0 \quad 1 \quad 0 \quad Lcos\left(\frac{\theta}{2}\right) \quad 0 \quad n + Lsin\left(\frac{\theta}{2}\right)\right) \tag{9}$$

Keeping in view the screw constraint topology graph from Fig.12, the screw constraint formulas for loops I-VIII are generated by applying the right handed rule. Loops I – III and loops V – VII are symmetric because these exhibit similar motion. Correspondingly, symmetric motion is observed for loop IV and loop VIII. Screw constraint formulas for loop I to loop VIII are given in equation (10) to equation (17), respectively.

**Loop-I:**
$$\begin{aligned} w_{27}S_{27} + w_{28}S_{28} + w_{29}S_{29} + w_{30}S_{30} \\ = w_{34}S_{34} + w_{33}S_{33} + w_{32}S_{32} + w_{31}S_{31} \end{aligned} \tag{10}$$

**Loop-II:**
$$\begin{aligned} w_{36}S_{36} + w_{35}S_{35} + w_{39}S_{39} + w_{40}S_{40} \\ = w_{41}S_{41} + w_{42}S_{42} + w_{43}S_{43} + w_{44}S_{44} \end{aligned} \tag{11}$$

**Loop-III:**
$$w_{25}S_{25} + w_{26}S_{26} + w_{34}S_{34} + w_{33}S_{33} \\ = w_{38}S_{38} + w_{37}S_{37} + w_{36}S_{36} + w_{35}S_{35} \quad (12)$$

**Loop-IV:**
$$w_{22}S_{22} + w_{21}S_{21} + w_{19}S_{19} + w_{20}S_{20} + w_2S_2 + w_1S_1 \\ = w_4S_4 + w_3S_3 + w_{28}S_{28} + w_{27}S_{27} + w_{26}S_{26} + w_{25}S_{25} \quad (13)$$

**Loop-V:**
$$w_{23}S_{23} + w_{24}S_{24} + w_{25}S_{25} + w_{26}S_{26} \\ = w_{17}S_{17} + w_{18}S_{18} + w_{21}S_{21} + w_{22}S_{22} \quad (14)$$

**Loop-VI:** $\quad w_{18}S_{18} + w_{17}S_{17} + w_8S_8 + w_7S_7 = w_6S_6 + w_5S_5 + w_{20}S_{20} + w_{19}S_{19} \quad (15)$

**Loop-VII:**
$$w_{14}S_{14} + w_{13}S_{13} + w_{12}S_{12} + w_{11}S_{11} \\ = w_{10}S_{10} + w_9S_9 + w_{15}S_{15} + w_{16}S_{16} \quad (16)$$

**Loop-VIII:**
$$w_{38}S_{38} + w_{37}S_{37} + w_{44}S_{44} + w_{43}S_{43} + w_{47}S_{47} + w_{48}S_{48} \\ = w_{23}S_{23} + w_{24}S_{24} + w_{14}S_{14} + w_{13}S_{13} + w_{45}S_{45} \\ + w_{46}S_{46} \quad (17)$$

Where $w_i$ are scalar angular velocities of the moving joints $i$. The DoF of the triple scissors links deployable modular unit is 1. The above system matrix is solved in MATLAB which indicates that the zero dimensions of above constraint matrix of the system is 1.

## 7. Application of Screw Theory: Analytical Kinematic Analysis

The kinematic analysis of the TSDAM is essential for understanding its transformation between stowed and deployed configurations. In its deployed state, the structure consists of a series of interconnected diamond-shaped units, each formed by straight bars arranged in a crisscross pattern. The transformation to the stowed form involves a coordinated folding mechanism at the pivot joints, allowing the diamond-shaped units to collapse inward. This process minimizes the overall dimensions of the structure, resulting in a compact, tightly packed form. The sequence of movements required for stowing begins with the rotation of the bars at each pivot joint, initiating the inward folding of the diamond units. As the bars rotate, the adjacent units follow a similar folding pattern, ensuring a synchronized and efficient collapse. This coordinated motion is driven by the geometric constraints imposed by the connected bars, which guide the structure towards a compact form with reduced width and depth. During this transition, the vertical alignment of the bars changes, with some bars rotating to lie parallel to the stowed axis, thus optimizing the storage space. Conversely, the deployment process involves reversing this sequence of movements. The pivot joints rotate outward, allowing the diamond-shaped units to extend and form the expanded lattice configuration. This process is characterized by the smooth unfolding of the bars, guided by the same geometric constraints that governed the stowing process. As each unit reaches its fully deployed position, the structure regains its original dimensions, with increased width and depth, suitable for various applications. The detailed kinematics based on screw constraint diagram, i.e., velocity and acceleration analysis, ensures that the structure can transition smoothly between stowed and deployed forms, retaining over all stability.

### 7.1 Velocity Analysis

The inner node O of a single module is designated as the fixed base. Moreover, its mobility in three dimensions stays constant. Fig. 12 shows the screw constraint diagram for the closed-loop unit mechanism. The velocity of each element of Loop III is given by equation (18).

$$\begin{cases} V_O = 0 \\ V_{M_1} = w_{25}S_{25} \\ V_M = w_{26}S_{26} \\ V_{N_1} = w_{36}S_{36} \\ V_N = w_{37}S_{37} \\ V_{M_2} = w_{34}S_{34} \\ V_{E_2} = w_{36}S_{36} \\ V_E = w_{33}S_{33} + w_{35}S_{35} \end{cases} \quad (18)$$

The velocity of every element is represented by the null vector $0_{6\times1}$. Equation (18) is used to get the screw velocities of the $M, M_2$ and $E$ node in Loop I. The velocities of every element in Loop I can be determined using the screw velocities derived in equation (19).

$$\begin{cases} V_M = w_{26}S_{26} \\ V_{M_2} = w_{34}S_{34} \\ V_E = w_{33}S_{33} + w_{35}S_{35} \\ V_{A_2} = V_M + w_{27}S_{27} \\ V_A = V_{A_2} + w_{28}S_{28} \\ V_{A_3} = V_A + w_{29}S_{29} \\ V_{D_1} = V_E + w_{32}S_{32} \\ V_D = V_{A_3} + w_{30}S_{30} \end{cases} \quad (19)$$

The same procedures can be applied to determine the screw velocity of every part in Loops II, IV-VIII. The angular velocity of each component as determined by screw velocity is given by equation (20).

$$w_i = w(V_i) \quad (20)$$

Where the initial three components of the screw velocity are denoted by $w(V_i)$. Equation (21) provides the mathematically calculates the center point linear velocity of every component.

$$v_i = v(V_i) + w(V_i) \times r_i \quad (21)$$

Where $v(V_i)$ represents the final three components of the velocity. Additionally, a vector called $r_i$ runs from the origin of the frame to the centroid point of component $i$. Referring to the left closed-loop unit mechanism and the coordinate system (CS) $O - XYZ$, the coordinate system $O - X_1Y_1Z_1$ is constructed at the same place as the right unit of mechanism, as seen in Fig. 12. The velocity of node $A$ in the coordinate framework $O - XYZ$ and node C in the coordinate framework $O - X_1Y_1Z_1$ are identical.

$$\begin{cases} {^O}v_A = {^O}v_C \\ {^O}w_A = {^O}w_C \end{cases} \quad (22)$$

The velocity relationship may be transferred to the entire triple scissors links deployable unit because the corresponding relationship between the nodes $A$ and $C$ in Fig. 12 is comparable to those of the other components. If the total number of closed-loop unit mechanisms in the triple scissor deployable unit is $N$, then $N$ coordinate systems $O - XYZ \sim O - X_{N-1}Y_{N-1}Z_{N-1}$ can be built. The global CS is set to $O - XYZ$, whereas the other CS are relative. The rotation angle between neighboring coordinate frames around the Z-axis is $\alpha$. The velocity of each component in the global CS $O - XYZ$ can be obtained from the expression in equation (23).

$$\begin{cases} {}^0v_i = {}^0_{0_j}R\,{}^{0_j}v_i \\ {}^0w_i = {}^0_{0_j}R\,{}^{0_j}w_i \end{cases} \quad (23)$$

Where $j$ is the number of CS and $i$ is the number of components. Equation (24) defines the rotation transformation matrix.

$$^0_{0_j}R = \begin{bmatrix} cos(j\alpha) & -sin(j\alpha) & 0 \\ sin(j\alpha) & cos(j\alpha) & 0 \\ 0 & 0 & 1 \end{bmatrix} \quad (24)$$

The aforementioned analysis allows for the derivation and expression of the angular and linear velocities at the center point of each component in the complete triple scissors links deployable module using coordinate transformation in the global CS.

## 7.2 Acceleration Analysis

By using the screw derivative to derive the generalized velocity of each component, the screw acceleration may be found. The screw acceleration is determined by applying the mathematical function given in equation (25).

$$A_i = \begin{bmatrix} \epsilon_i \\ a - w \times v \end{bmatrix} \quad (25)$$

Where $a$ is the linear acceleration at the centroid of component $i$, $A_i$ is the screw acceleration of component $i$ in the triple scissor deployable unit, and $\epsilon_i$ is the angular acceleration of the coincidence point with respect to the origin of the component $i$ reference CS. Using the screw acceleration, by the Lie operation [], one can calculate the acceleration of the inner node E in Loop III of the screw constraint diagram, Fig. 12.

$$\begin{cases} \epsilon_{25}S_{25} + \epsilon_{26}S_{26} + \epsilon_{34}S_{34} + \epsilon_{33}S_{33} + S^1_{Lie} = O^\epsilon E \\ \epsilon_{38}S_{38} + \epsilon_{37}S_{37} + \epsilon_{36}S_{36} + \epsilon_{35}S_{35} + S^2_{Lie} = O^\epsilon E \end{cases} \quad (26)$$

Where $O^\epsilon E$ is angular acceleration of node $E$ relative to the node $O$. The two Lie [] expressions are arranged as mentioned in equation (27).

$$\begin{cases} S^1_{Lie} = Lie[w_{25}S_{25}, w_{26}S_{26}] + Lie[w_{25}S_{25}, w_{34}S_{34}] + Lie[w_{25}S_{25}, w_{33}S_{33}] + \\ \qquad Lie[w_{26}S_{26}, w_{34}S_{34}] + Lie[w_{26}S_{26}, w_{33}S_{33}] + Lie[w_{34}S_{34}, w_{33}S_{33}] \\ S^2_{Lie} = Lie[w_{38}S_{38}, w_{37}S_{37}] + Lie[w_{38}S_{38}, w_{36}S_{36}] + Lie[w_{38}S_{38}, w_{35}S_{35}] + \\ \qquad Lie[w_{37}S_{37}, w_{36}S_{36}] + Lie[w_{37}S_{37}, w_{35}S_{35}] + Lie[w_{36}S_{36}, w_{35}S_{35}] \end{cases} \quad (27)$$

The Lie [] is the representation for Lie brackets which satisfy anti-symmetry property as well as Jacobi identity property. For each term $Lie[w_iS_i, w_jS_j] = w_iw_j[S_i, S_j]$ where $[S_i, S_j] = \sum_k \epsilon^k_{ij} S_k$.

By combining equations (26) and (27) we get;

$$\begin{aligned} \epsilon_{25}S_{25} + \epsilon_{26}S_{26} + \epsilon_{34}S_{34} + \epsilon_{33}S_{33} + S^1_{Lie} \\ = \epsilon_{38}S_{38} + \epsilon_{37}S_{37} + \epsilon_{36}S_{36} + \epsilon_{35}S_{35} + S^2_{Lie} \end{aligned} \quad (28)$$

Similar steps can be followed to generate the acceleration screw constraints equations of Loop I, II, and loop IV-VIII in Fig. 12. Screw constraint equations can be used to determine the screw acceleration of each component given the input angular acceleration $\epsilon_{25}$.

Equation (25) can be used to determine the angular as well as linear accelerations of centroid of each component once the screw acceleration has been determined.

$$\begin{cases} \epsilon_i = \epsilon(\epsilon_i S_i) \\ a_i = a(a(\epsilon_i S_i) + w_i \times v_i) + [\epsilon(\epsilon_i S_i) \times r_i] + [w_i \times (w_i \times r_i)] \end{cases} \quad (29)$$

The first three elements in equation (29) indicate the origin of the screw acceleration represented by $\epsilon(\epsilon_i S_i)$, and the final three elements are acceleration, illustrated by $a(\epsilon_i S_i)$. Similar to the velocity analysis, every part in each unit mechanism with identical position has identical acceleration in its individual CS. This can be displayed in the global coordinates system in the form of mathematical expression given in equation (30).

$$\begin{cases} O_{\epsilon_i} = {}^0_{0_j}R^{0_j}\epsilon_i \\ O_{a_i} = {}^0_{0_j}R^{0_j}a_i \end{cases} \quad (30)$$

The angular and centroid line accelerations of every part of the triple scissors deployable unit mechanism can be computed in the global CS by applying the above method.

## 8. Results and Discussion
### 8.1 Computational vs Virtual Experiments Results

The analytical model, for different kinematic parameters that define the movement and behavior of the deployable module of TSDAM during deployment, is developed by applying the screw theory approach in section 7. This model is required to be validated and assessed, for consistency in motion pattern and results, through simulations analysis in SolidWorks and MATLAB. The virtual experiments provide an estimate of the kinematic analysis for the complete mechanism as it changes its configuration from the stowed to the completely deployed state. Every deployable unit of TSDAM comprises of several components and each type of components is placed in the exact same position within every modular unit. The mechanism with its local CS aligned with the global CS is preferred for investigation to simplify the explanation of these procedures. Considering the unit mechanism in Fig. 12, the global CS is established at node A, and nodes B and D are selected for simulation analysis and verification of theoretical model. Kinematic simulations are performed on a chosen links of the mechanism to illustrate the coordinated motion of each unit module as it contains symmetry in the system as a whole. Fig. 13 (a-b) and Fig. 14 (a-b) shows the comparative analysis for linear velocity, angular velocity, linear acceleration, and angular acceleration, respectively. The comparison demonstrates excellent agreement between the analytical and simulation models, with minimal discrepancies, confirming the accuracy and reliability of both approaches. The results reveal consistent trends across the kinematic parameters. Linear velocity, as shown in Fig. 13(a), increases steadily, indicating a uniform deployment process. Angular velocity trends in Fig. 13(b) demonstrate synchronized rotational motion, further confirming the coordinated behavior of the system. Linear acceleration (Fig. 14(a)) remains stable overall, with slight deviations, while angular acceleration (Fig. 14(b)) shows controlled oscillations. This alignment between analytical and simulation results validates the reliability of the screw theory-based model and its successful implementation.

The kinematic analysis of specific nodes (A, B, and D), in Fig. 13(a-b), highlights the velocity behavior of the system during deployment. Node A achieves the highest linear velocity, peaking at approximately 360mm/s. This occurs because Node A, positioned at the tip of the deployment path, experiences the largest displacement as the scissor links extend. This demonstrates the effectiveness of the deployment to achieve quick extension. Node B reaches a moderate velocity of 150mm/s due to its intermediate position, experiencing partial displacement. Node D, located near the fixed origin, exhibits the lowest velocity, remaining under 100mm/s, as its motion is more constrained by its proximity to the base. Similarly, angular velocity trends reflect similar behavior that indicates rotational motion executed by different links of antenna. Node A achieves the highest angular velocity, around 4rad/s, due to amplified rotational effects at the extended linkages. Node B stabilizes at 2rad/s, while Node D, with minimal rotational motion, remains below 1rad/s. These trends highlight the effective design of the antenna, ensuring maximum velocity at the distal nodes for rapid deployment while maintaining controlled motion closer to the base for structural stability.

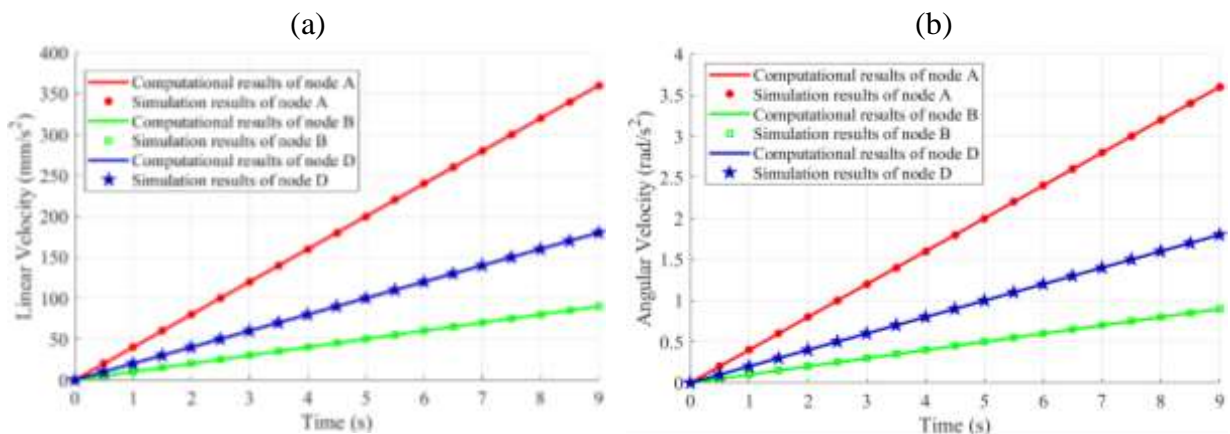

**Fig. 13.** Comparative analysis to validate the analytical and simulation results at selected nodes of deployable antenna modular unit (a) Linear velocity (b) Angular velocity

Acceleration response in Fig. 14 (a-b) provide further insights into the forces acting on the deployment system of deployable modular unit. Node A exhibits the highest linear acceleration, stabilizing around 50mm/s², reflecting the rapid extension forces acting at the tip. Node B stabilizes at 25mm/s², while Node D, constrained by its proximity to the fixed origin, experiences the lowest acceleration at approximately 10mm/s². Angular acceleration follows a similar pattern ensuring that the antenna is stable and rotationally controlled. The Node A stabilizes at 0.5rad/s², Node B at 0.3rad/s², and Node D at 0.1rad/s². These results confirm that the deployment forces are effectively distributed, with higher accelerations at distal nodes ensuring efficient motion and lower accelerations near the base for stability.

Overall, the results in Fig. 13 and Fig. 14 confirm the accuracy of the analytical and simulation models in capturing the kinematic behavior of the deployable triple scissors link antenna. The consistent trends in velocity and acceleration profiles, with minimal deviations, validate the antenna's design for rapid, reliable, and synchronized deployment, making it a robust and effective solution for deployable structures.

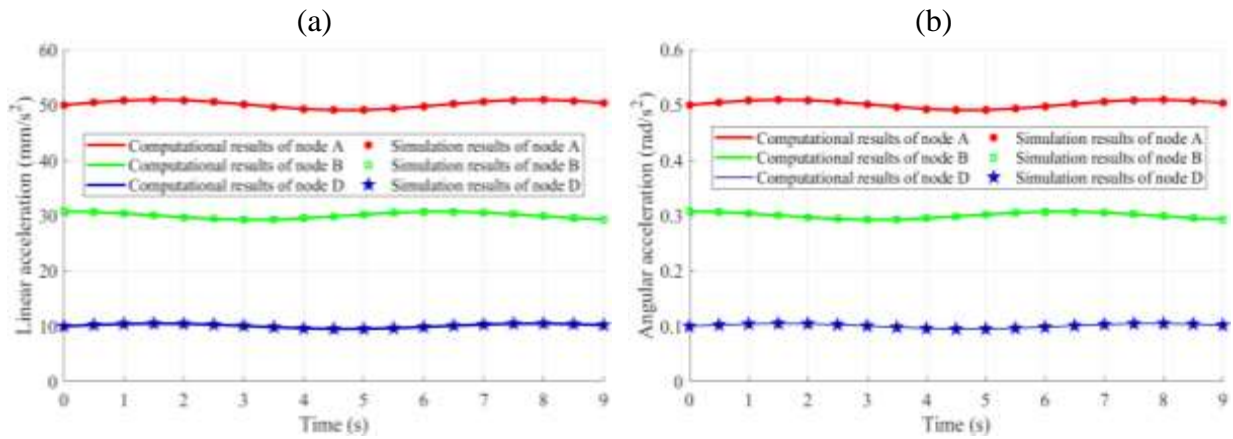

**Fig. 14.** Comparative analysis to validate the analytical and simulation results at selected nodes of deployable antenna modular unit (a) Linear acceleration (b) Angular acceleration

Fig. 15 (a-b) and Fig. 16 (a-b) illustrates the variations in maximum, minimum and mean values of linear and angular velocities and accelerations at different node positions of a single modular unit of the TSDAM, respectively. Nodes at the periphery (e.g., **A, G, C, I**) exhibit higher values for linear and angular velocities and accelerations due to their greater displacement and active role in extending the system.

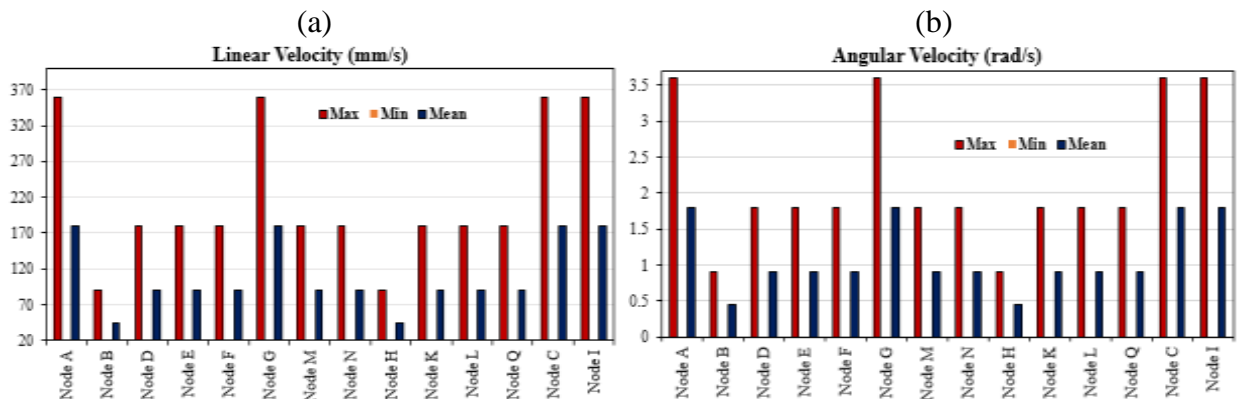

**Fig. 15.** Analysis of Maximum, Minimum and Mean velocities response at different nodal positions of triple scissors links modular unit (a) linear velocities (b) angular velocities

At these locations amplification in motion is observed due to the extension of scissors mechanism. Nodes closer to the center (e.g., **M**, **K**, **L**) show lower values of linear velocities. These nodes remain more constrained as they are closer to the base of the modular unit, which restricts their displacement. Additionally, the mean values provide a balanced perspective, reflecting the overall uniformity of the system in velocity distribution. Moreover, the angular velocity and acceleration values are consistent with the geometry of the scissor links, ensuring coordinated rotational motion across the modular unit.

Analysis in Fig. 16 (a-b) show that the outer nodes (e.g., **A**, **B**) exhibit higher linear and angular accelerations, reflecting the significant forces acting on them to achieve rapid deployment with amplified rotational motion. These nodes are responsible for extending the system outward, requiring higher acceleration to overcome resistance and maintain smooth motion. The differences in these kinematic values are primarily influenced by the geometric position of the nodes and their roles within the mechanism. Peripheral nodes actively contribute to the deployment process, requiring higher velocity and acceleration, whereas central nodes serve as stabilizing pivots or intermediate points. The screw graph representation helps explain the distribution of motion across the mechanism, ensuring coordinated and efficient deployment. Additionally, the kinematic analysis validates the efficient design of the TSDAM, ensuring rapid deployment with controlled motion. Higher kinematic values at outer nodes enable effective deployment. The lower values at central nodes maintain structural stability, prevent excessive vibrations, and contribute to validate the balanced and reliable performance of the modular unit.

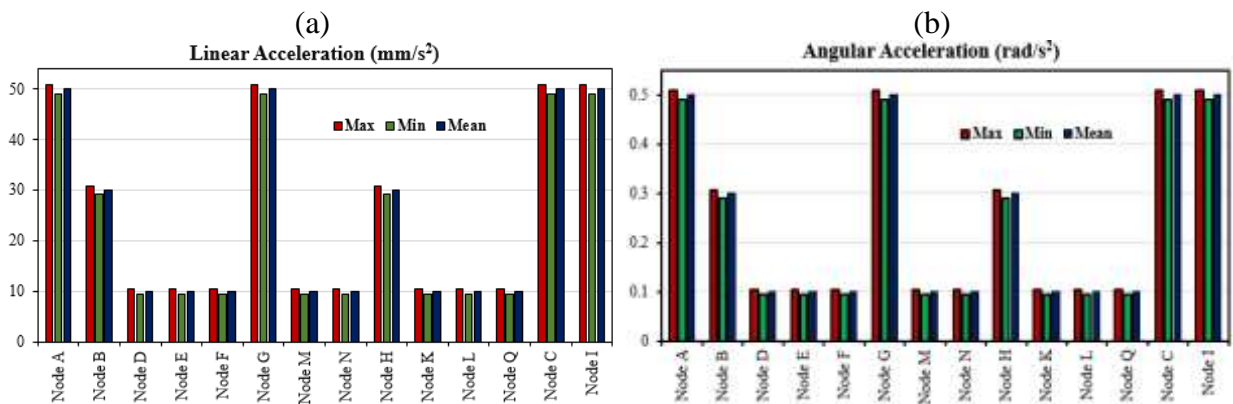

**Fig. 16.** Analysis of Maximum, Minimum and Mean acceleration response at different nodal positions of triple scissors links modular unit (a) linear acceleration (b) angular acceleration

**8.2 Analysis of Virtual Experiments results on two adjacent modular units**

Fig. 17 illustrates the two adjacent modular units of the deployable antenna used for simulation analysis, while Fig. 18 presents the kinematic parameters i.e., linear and angular velocity and acceleration, simulated in SolidWorks.

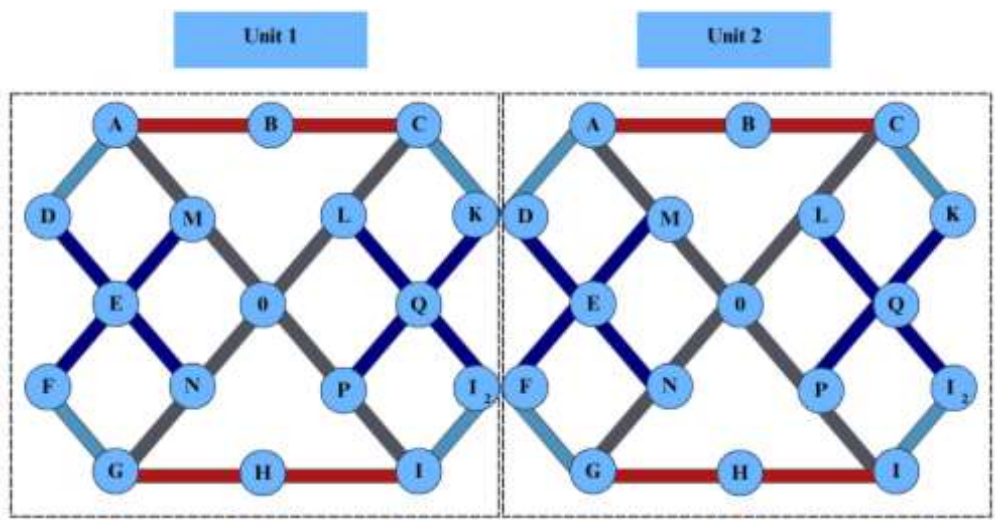

**Fig. 17.** Schematic representation of two adjacent deployable modular units.

The motion pattern of these units clearly depicts the consistency and synchronization between their kinematic responses during operation. Comparing the velocity profiles of Units 1 and 2 reveals a significant similarity, with minimal deviations observed throughout the simulation time. The mean squared error (MSE) between the velocity profiles is approximately 0.002775, and the root mean square error (RMSE) is 0.05267, indicating that the magnitude of deviations is small and ensures aligned velocity profiles between the two units.

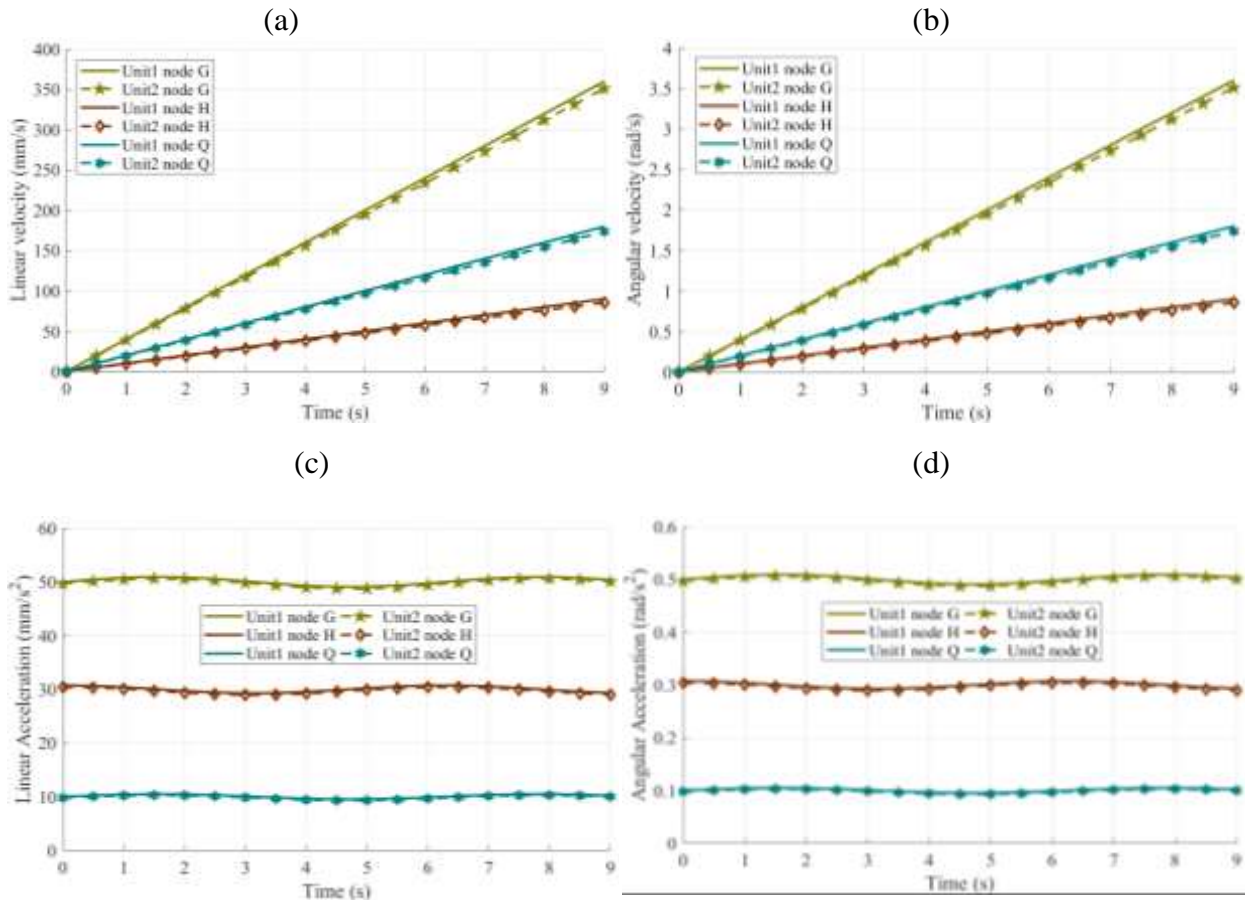

**Fig. 18.** Comparison of simulation results on selected nodes of two adjacent modular units (a) Linear velocity (b) Angular velocity (c) Linear acceleration (d) Angular acceleration

Similarly, the acceleration profiles are strongly correlated, with the MSE for acceleration being $1.089 \times 10^{-5}$ and the RMSE approximately 0.0033, highlighting negligible differences in acceleration between the units. These small error values in both velocity and acceleration metrics show the accuracy and reliability of the simulated model in SolidWorks. The detailed trends for each parameter further validate these findings. Linear velocity (Fig. 18a) exhibits a steady increase across all nodes, reflecting the smooth deployment mechanism of the adjacent modular units. Angular velocity (Fig. 18b) follows a consistent trend, with variations attributed to the kinematic configuration and rotational dynamics of the links during deployment. Linear acceleration (Fig. 18c) remains stable with minimal fluctuations, while the angular acceleration (Fig. 18d) shows controlled variations, further reinforcing the precision of the simulated model. These small error values and stable kinematic profiles collectively underscore the accuracy, reliability, smooth and coordinated motion of the adjacent deployable modular units during the simulation process.

Table 6 and Fig. 19 show the comparative simulation analysis of maximum, minimum and average linear and angular velocities and accelerations at selected node points on adjacent modular units of TSDAM. The virtual experiments performed in SolidWorks show the symmetric nature of the model. It can be observed that the linear and angular kinematic response of the two adjacent modular units is highly comparable to each other with very minor deviations. This uniformity ensures the robust kinematic design of the TSDAM.

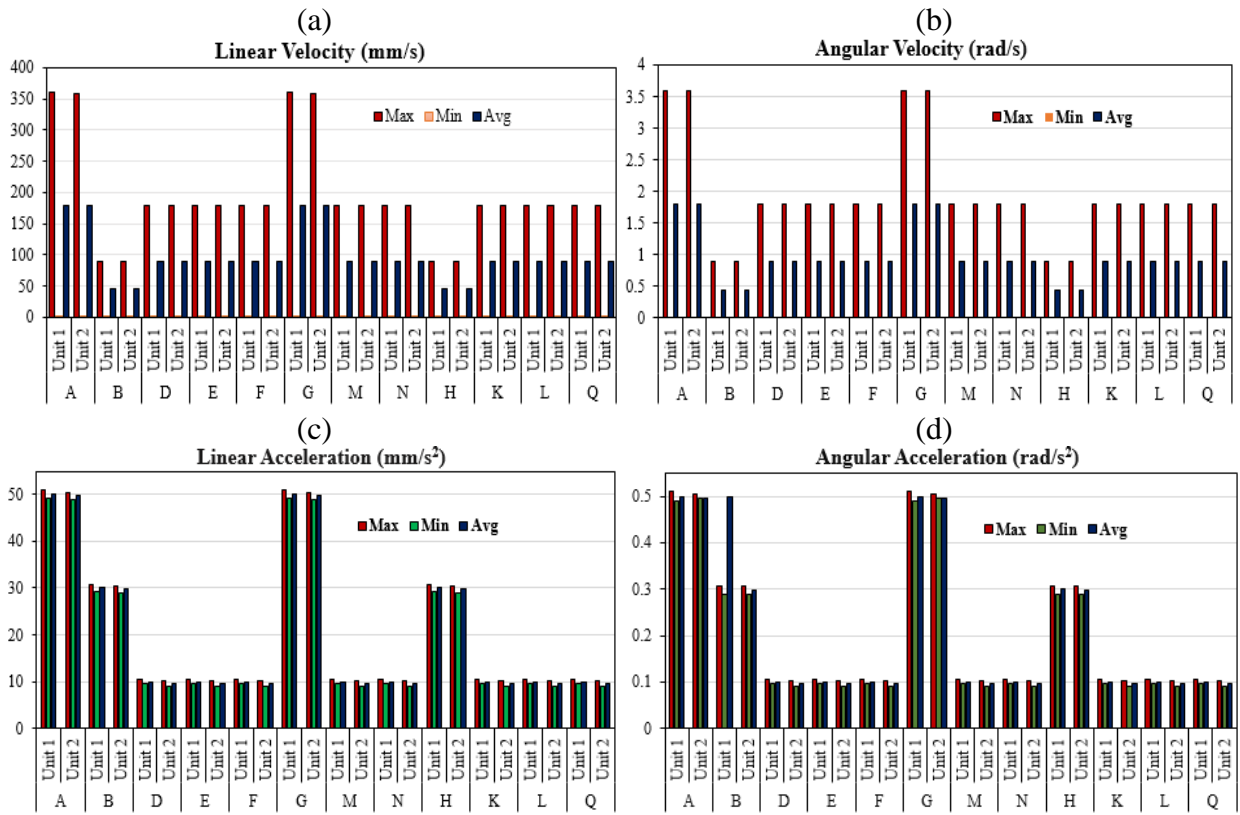

**Fig. 19.** Comparative **a**nalysis of Maximum, Minimum and Average linear and angular velocities and accelerations at selected node points on adjacent modular units of TSDAM

The closely-identical, maximum, average and minimum, values of linear and angular motion between adjacent deployable modules in Table 6 confirm coordinated motion that ensures the proposed geometry and its deployment functions. The minor variances in the responses at different node positions of Unit 1 and Unit 2 can be attributed to the distribution of forces during the quick deployment phase that typically arise during the virtual experiments. Generally, the synchronized results support the deployable geometry, kinematic design and the stability, reliability, and suitability of simulation approach for analysis of deployable space antenna. Moreover, the uniform transfer of motion between adjacent modular units of TSDAM guarantees consistency of successful structural deployment that is critical parameter to investigate the performance of antenna.

**Table 6.** Minimum, Maximum and average values of velocity and acceleration at selected node positions of two adjacent modular units

| Nodes | Unit | Linear Velocity (mm/s) | | | Angular Velocity (rad/s) | | | Linear Acceleration (mm/s$^2$) | | | Angular Acceleration (rad/s$^2$) | | |
|---|---|---|---|---|---|---|---|---|---|---|---|---|---|
| | | Max | Min | Avg | Max | Min | Avg | Max | Min | Avg | Max | Min | Avg |
| A | 1 | 360 | 1 | 180 | 3.6 | 0.01 | 1.8 | 50.94 | 49.07 | 50 | 0.5094 | 0.49 | 0.5 |
|   | 2 | 358.7 | 1 | 179.35 | 3.58 | 0.01 | 1.79 | 50.35 | 49.01 | 49.68 | 0.5035 | 0.496 | 0.496 |
| **B** | 1 | 90 | 1 | 45 | 0.9 | 0.01 | 0.45 | 30.75 | 29.25 | 30 | 0.3075 | 0.29 | 0.5 |
|   | 2 | 89.13 | 1 | 44.46 | 0.89 | 0.01 | 0.44 | 30.51 | 29.03 | 29.87 | 0.3051 | 0.290 | 0.298 |
| **D** | 1 | 180 | 1 | 90 | 1.8 | 0.01 | 0.9 | 10.49 | 9.51 | 10 | 0.1049 | 0.095 | 0.1 |
|   | 2 | 179.3 | 1 | 89.46 | 1.79 | 0.01 | 0.89 | 10.13 | 9.07 | 9.7 | 0.1013 | 0.090 | 0.097 |
| **E** | 1 | 180 | 1 | 90 | 1.8 | 0.01 | 0.9 | 10.49 | 9.51 | 10 | 0.1049 | 0.095 | 0.1 |
|   | 2 | 179.3 | 1 | 89.46 | 1.79 | 0.01 | 0.89 | 10.13 | 9.07 | 9.7 | 0.1013 | 0.090 | 0.097 |
| **F** | 1 | 180 | 1 | 90 | 1.8 | 0.01 | 0.9 | 10.49 | 9.51 | 10 | 0.1049 | 0.095 | 0.1 |
|   | 2 | 179.3 | 1 | 89.46 | 1.79 | 0.01 | 0.89 | 10.13 | 9.07 | 9.7 | 0.1013 | 0.090 | 0.097 |
| **G** | 1 | 360 | 1 | 180 | 3.6 | 0.01 | 1.8 | 50.94 | 49.07 | 50 | 0.5094 | 0.49 | 0.5 |
|   | 2 | 358.7 | 1 | 179.35 | 3.58 | 0.01 | 1.79 | 50.35 | 49.01 | 49.68 | 0.5035 | 0.496 | 0.496 |
| **M** | 1 | 180 | 1 | 90 | 1.8 | 0.01 | 0.9 | 10.49 | 9.51 | 10 | 0.1049 | 0.095 | 0.1 |
|   | 2 | 179.3 | 1 | 89.46 | 1.79 | 0.01 | 0.89 | 10.13 | 9.07 | 9.7 | 0.1013 | 0.090 | 0.097 |
| **N** | 1 | 180 | 1 | 90 | 1.8 | 0.01 | 0.9 | 10.49 | 9.51 | 10 | 0.1049 | 0.095 | 0.1 |
|   | 2 | 179.3 | 1 | 89.46 | 1.79 | 0.01 | 0.89 | 10.13 | 9.07 | 9.7 | 0.1013 | 0.090 | 0.097 |
| **H** | 1 | 90 | 1 | 45 | 0.9 | 0.01 | 0.45 | 30.75 | 29.25 | 30 | 0.3075 | 0.29 | 0.3 |
|   | 2 | 89.13 | 1 | 44.46 | 0.89 | 0.01 | 0.44 | 30.51 | 29.03 | 29.87 | 0.3051 | 0.290 | 0.298 |
| **K** | 1 | 180 | 1 | 90 | 1.8 | 0.01 | 0.9 | 10.49 | 9.51 | 10 | 0.1049 | 0.095 | 0.1 |
|   | 2 | 179.3 | 1 | 89.46 | 1.79 | 0.01 | 0.89 | 10.13 | 9.07 | 9.7 | 0.1013 | 0.090 | 0.097 |
| **L** | 1 | 180 | 1 | 90 | 1.8 | 0.01 | 0.9 | 10.49 | 9.51 | 10 | 0.1049 | 0.095 | 0.1 |
|   | 2 | 179.3 | 1 | 89.46 | 1.79 | 0.01 | 0.89 | 10.13 | 9.07 | 9.7 | 0.1013 | 0.090 | 0.097 |
| **Q** | 1 | 180 | 1 | 90 | 1.8 | 0.01 | 0.9 | 10.49 | 9.51 | 10 | 0.1049 | 0.095 | 0.1 |
|   | 2 | 179.3 | 1 | 89.46 | 1.79 | 0.01 | 0.89 | 10.13 | 9.07 | 9.7 | 0.1013 | 0.090 | 0.097 |

## 9. Conclusion

This study proposes a novel **Triple Scissors Deployable Antenna Mechanism (TSDAM)**, designed to overcome critical challenges associated with large-aperture space antennas by ensuring high precision, structural reliability, and efficient deployment. The motion relationship between the components was effectively described using the screw constraint diagram, and the DoF of the mechanism was determined through matrix-based operations. Leveraging screw theory for analytical kinematic analysis, the performance of mechanism was rigorously validated through virtual experiments conducted in SolidWorks and MATLAB. The single DoF mechanism demonstrated exceptional synchronization and smooth deployment, offering a highly reliable solution for space applications. The motion relationships were effectively captured using screw constraint diagrams, and the DoF was analytically determined through matrix-based operations. Its integration with the properties of 7075-T6 aluminum enhances the real-world applicability of the structural design and opens doors to eventual prototyping and terrestrial validation. The optimized 12-units configuration proved to be the most effective, achieving a maximum deformation of just 0.01048mm**, a** storage ratio of 15.3 for height and volume**,** and a deployment time of 53 seconds**.** Although the 24-units configuration exhibited superior stowage compactness, its comparatively higher deformation limited its structural stability, confirming the 12-units configuration as the optimal choice for demanding space missions. These findings validate the robustness, feasibility, and practicality of the TSDAM for large-scale deployable structures in space environments.

In conclusion, this study establishes a robust design framework for deployable antennas, offering an optimal balance between structural integrity, deployment efficiency, and stowage compactness. The TSDAM demonstrates significant advancements in the design and analysis of large-aperture deployable mechanisms, serving as a valuable reference for future space structures. To further enhance the performance of mechanism, lightweight composite materials should be explored for structural optimization. Experimental validation under space-simulated conditions is essential to ensure reliability. Moreover, incorporating advanced control systems and scalability studies will enable greater adaptability and precision for diverse space missions.